\documentstyle[epsf]{mn}

\newif\ifAMStwofonts

\newcommand{\phl} {$11.9~\lambda$ }
\newcommand{\hrs} {$16.7~\lambda$ }

\ifoldfss
  \ifCUPmtlplainloaded \else
    \NewTextAlphabet{textbfit} {cmbxti10} {}
    \NewTextAlphabet{textbfss} {cmssbx10} {}
    \NewMathAlphabet{mathbfit} {cmbxti10} {} 
    \NewMathAlphabet{mathbfss} {cmssbx10} {} 
  \fi
  \ifAMStwofonts
    \ifCUPmtlplainloaded \else
      \NewSymbolFont{upmath} {eurm10}
      \NewSymbolFont{AMSa} {msam10}
      \NewMathSymbol{\upi}     {0}{upmath}{19}
      \NewMathSymbol{\umu}     {0}{upmath}{16}
      \NewMathSymbol{\upartial}{0}{upmath}{40}
      \NewMathSymbol{\leqslant}{3}{AMSa}{36}
      \NewMathSymbol{\geqslant}{3}{AMSa}{3E}

    \fi
  \fi
\fi 

\ifnfssone
  \newmathalphabet{\mathit}
  \addtoversion{normal}{\mathit}{cmr}{m}{it}
  \addtoversion{bold}{\mathit}{cmr}{bx}{it}
  \newmathalphabet{\mathbfit} 
  \addtoversion{normal}{\mathbfit}{cmr}{bx}{it}
  \addtoversion{bold}{\mathbfit}{cmr}{bx}{it}
  \newmathalphabet{\mathbfss} 
  \addtoversion{normal}{\mathbfss}{cmss}{bx}{n}
  \addtoversion{bold}{\mathbfss}{cmss}{bx}{n}
  \ifAMStwofonts
    \ifCUPmtlplainloaded \else
      \UseAMStwoboldmath
      \makeatletter
      \new@mathgroup\upmath@group
      \define@mathgroup\mv@normal\upmath@group{eur}{m}{n}
      \define@mathgroup\mv@bold\upmath@group{eur}{b}{n}
      \edef\UPM{\hexnumber\upmath@group}
      \new@mathgroup\amsa@group
      \define@mathgroup\mv@normal\amsa@group{msa}{m}{n}
      \define@mathgroup\mv@bold\amsa@group{msa}{m}{n}
      \edef\AMSa{\hexnumber\amsa@group}
      \makeatother
      \mathchardef\upi="0\UPM19
      \mathchardef\umu="0\UPM16
      \mathchardef\upartial="0\UPM40
      \mathchardef\leqslant="3\AMSa36
      \mathchardef\geqslant="3\AMSa3E
    \fi
  \fi
\fi 

\ifnfsstwo
  \DeclareMathAlphabet{\mathbfit}{OT1}{cmr}{bx}{it}
  \SetMathAlphabet\mathbfit{bold}{OT1}{cmr}{bx}{it}
  \DeclareMathAlphabet{\mathbfss}{OT1}{cmss}{bx}{n}
  \SetMathAlphabet\mathbfss{bold}{OT1}{cmss}{bx}{n}
  \ifAMStwofonts
    \ifCUPmtlplainloaded \else
      \DeclareSymbolFont{UPM}{U}{eur}{m}{n}
      \SetSymbolFont{UPM}{bold}{U}{eur}{b}{n}
      \DeclareSymbolFont{AMSa}{U}{msa}{m}{n}
      \DeclareMathSymbol{\upi}{0}{UPM}{"19}
      \DeclareMathSymbol{\umu}{0}{UPM}{"16}
      \DeclareMathSymbol{\upartial}{0}{UPM}{"40}
      \DeclareMathSymbol{\leqslant}{3}{AMSa}{"36}
      \DeclareMathSymbol{\geqslant}{3}{AMSa}{"3E}
    \fi
  \fi
\fi 

\ifCUPmtlplainloaded \else
  \ifAMStwofonts \else 
    \def\upi{\pi}
    \def\umu{\mu}
    \def\upartial{\partial}
  \fi
\fi

\title{A 33 GHz interferometer for CMB observations on Tenerife}
\author[S.J.Melhuish et al.]
  {S.J. Melhuish,$^1$\thanks{E-mail: sjm@jb.man.ac.uk} S.Dicker,$^1$ R.D. Davies,$^1$ 
  C.M. Guti\'{e}rrez,$^2$ R.A. Watson,$^1$ \newauthor R.J. Davis,$^1$ R. Hoyland$^2$ and R. Rebolo$^2$\\
 $^1$University of Manchester, Nuffield Radio Astronomy Laboratories,
Jodrell Bank, Macclesfield, Cheshire,  SK11 9DL\\
 $^2$Instituto de Astrofisica de Canarias, 38200 La Laguna, Tenerife, Spain}
\date{Accepted 1999 January 5. Received 1998 December 30; in original form 1997 December 15}

\pagerange{\pageref{firstpage}--\pageref{lastpage}}
\pubyear{1998}

\begin{document}

\maketitle

\label{firstpage}

\begin{abstract}

We describe a new high sensitivity experiment for observing  cosmic
microwave background (CMB) anisotropies.  The instrument is a  2-element
interferometer operating at 33~GHz with a $\sim$3~GHz bandwidth.   It is
installed on the high and dry Teide Observatory site on Tenerife where 
 successful beam-switching observations have been made at this frequency.  
Two realizations of the interferometer have been tested with element
separations of \phl and \hrs.
  The resulting angular resolution of $\sim2^\circ$ was 
chosen to explore the amplitude of CMB structure on the large angular scale 
side of the Doppler (acoustic)
 peak.  It is found that observations are unaffected by water vapour for 
more than 70 per cent of the time when the sensitivity is limited 
by the receiver noise alone.  Observations over several months are expected 
to give an rms noise level of $\sim 10 - 20~\mu$K covering $\sim$100 
resolution elements.  
Preliminary results show stable operation of the interferometer with the 
detection of discrete radio sources as well as the Galactic plane at Dec =
+41\degr~ and $-$29\degr.

\end{abstract}


\section{Introduction}
\label{sect:introduction}

The detection of angular structure in the cosmic microwave background (CMB) 
gives a wealth of cosmological data.  Observations in the angular range 
100\degr~ to 0\fdg1 or less can constrain such cosmological parameters as 
$\Omega_0$, $\Omega_B$, $H_0$ and the spatial spectral index $n$.  On the 
larger angular scales (say $>$5\degr) the tensor component contribution can be 
quantified relative to the scalar Sachs--Wolfe contribution.  Advances in 
this field require a good coverage of the angular spectrum at high 
sensitivity.

Our experiments on the 2400~m high Teide Observatory site on Tenerife have 
been of the beam-switching type and covered the angular range $\sim$4\degr~ to 
16\degr~ by switching a 5\degr~ beam through angles of $\pm$8\degr~
\cite{d1996}.
  Through the use of scaled experiments at 10, 15 and 33 GHz, we 
can separate the Galactic and CMB components \cite[1997]{h1994}.  
These experiments have provided the first detection of individual hot and 
cold features which have now been confirmed by an independent observing 
group \cite{l1995,g1997}. 

We describe here a new high sensitivity interferometer system, also 
located at Teide Observatory, which can investigate structure on 2\degr~ 
angular scales. The separation of the horn elements can be increased to
explore angular scales as small as 1\degr. With longer baselines the
effective ``filling factor'' would be too small for sensitive measurements.
The 1 to 2\degr~ range of angles
 is on the large scale side of the first of the  Doppler (acoustic) peaks 
produced by
photon and matter sources which lead to a maximum amplitude at around the
horizon scale at last scattering \cite{wss1994}. The interferometer will be used
to  
measure the slope of the CMB spatial spectrum over this range.
A frequency of 33~GHz was chosen, a frequency at which we have already 
shown that the Galactic component $\Delta T_{\rm rms} < 5\;\mu$K \cite{h1997} 
at $5\degr - 15\degr~$ angular resolution. On angular scales of $1\degr -
2\degr$ this will be significantly less \cite{l1996,v1997}. 
The corresponding  CMB $\Delta
T_{\rm rms}$ is expected to be $50 - 100~\mu$K.   Furthermore, a particular
advantage of an interferometer is its much  stronger rejection of atmospheric
fluctuations compared with a  beam-switching system \cite{w1994,c1995}.
Meteorolgical features subtending angles larger than the interferometer lobe
size do not produce a large correlator output signal. However, even smaller
features can be supressed if there is some east--west motion of the ``cloudlets''
through the beam. During an integration the correlator output signal rotates in
phase, effecting a cancelation. The cloudlets appear stretched in the direction
of motion, and with even a small east--west component this tends to quench the
interferometer response.

In this paper  Section~\ref{sect:desc}
 describes the design concepts 
and realization of the interferometer while in Section~\ref{sect:performance}
 we assess the 
performance of the instrument. Section~\ref{sect:atmos} contains information
on  the ability of the interferometer to reject atmospheric effects, and 
Section~\ref{sect:astro} describes astronomical  commissioning observations
with two configurations of the interferometer.   The future astronomical
programme is outlined in Section~\ref{sect:conc}.

\section{Description of the interferometer system }
\label{sect:desc}

The interferometer system was built and tested at Jodrell Bank and then 
moved to Teide Observatory, Iza\~{n}a, for observations.  This observing site 
has low levels of precipitable water vapour (pwv $<$ 2 mm) for much of the 
year and permits the collection of high quality data from the 33 GHz
beam-switching radiometer  for $\sim$20 per cent of the time.  It was
anticipated that an interferometer
 at this frequency would be usable for a much higher fraction of the 
time;  our experience confirms this and is quantified in
Section~\ref{sect:atmos}.

High sensitivity to CMB anisotropy 
can be achieved by using cryogenically cooled 
low noise receiver systems operating with large ($\sim$10 per cent) 
bandwidths.  
The required sensitivity of $\sim$10~$\mu$K per resolution element is 
obtained by repeated observations of the chosen area of sky.  The design 
of the interferometer described below owes much to the experience 
accumulated with the 5 GHz wide bandwidth interferometer operated at 
Jodrell Bank \cite{m1997}.

\subsection{Concept of the CMB interferometer}
\label{sect:concept}

\begin{figure}
\begin{center}
\leavevmode\epsfbox{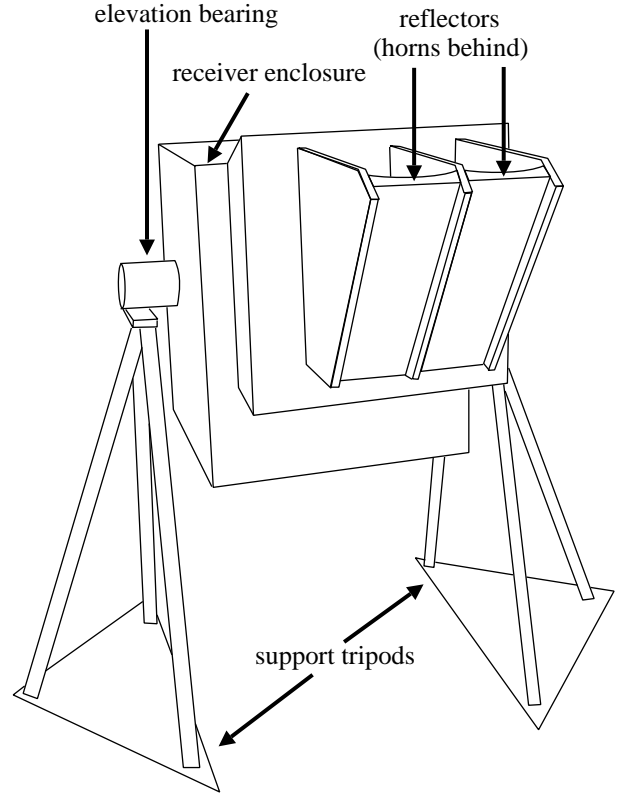}
\end{center}
\caption{\label{phys}
A three dimensional representation of the interferometer in the
\hrs configuration. A weather--sealed box contains the cooled
receivers and the correlator, with the horn-reflector 
antenna system projecting from the
front face. The assembly may be rotated in the elevation direction upon
its support tripods. }
\end{figure}

The interferometer consists of two 
horn apertures fixed in an east--west line.  
Earth rotation then sweeps the interferometer fringe pattern across the 
sky giving 24${\rm ^h}$~ coverage in right ascension (RA) each day.  The 
declination of observation 
is set by tilting the interferometer assembly in the meridian
 plane to the appropriate elevation.  The physical arrangement is shown in 
Fig.~\ref{phys}.  The aim is to make a high sensitivity map of the band 
of sky, centred on Dec = +41\degr, that has been explored on somewhat larger 
angular scales at 5, 10, 15 and 33 GHz \cite{m1997,g1995,g1997,h1997}.

The operating frequency was chosen to be 33 GHz where our previous 
experience with beam-switching radiometers on $5\degr - 15\degr$ angular scales
indicated a level of Galactic contamination less than 
10 per cent of the intrinsic CMB fluctuation amplitude.
Since the interferometer is sensitive to structures with smaller angular scales,
it is expected that any Galactic contribution will be less still,
because of the downwards slope of the Galactic spatial power spectrum.
On the 2\degr~angular scale of this interferometer both the synchrotron
\cite{l1996} and the free--free \cite{v1997} Galactic components will be a
factor of 2 -- 4 less than found in our Tenerife beam-switching observations.
Moreover, at this frequency the atmospheric contamination of the signals 
observed with the interferometer will be substantially lower than that 
experienced with the 33~GHz beam-switching radiometer at Teide Observatory.  
The system was designed to have a 3~GHz (10 per cent) bandwidth which can 
be readily realized using analogue technology for the correlator.

The design aim was to achieve a resolution of $2\degr\times2\degr$.
 To obtain the highest brightness temperature sensitivity, 
the receiving horns were placed adjacent to one another leaving only a small
$\sim$10~mm gap.  By choosing a primary beam of $5\degr\times2\degr$ 
 elongated in the EW direction, a synthesized beam measuring approximately 
$2\degr\times2\degr$ can be produced by the interferometer response.

\begin{figure}
\begin{center}
\leavevmode\epsfbox{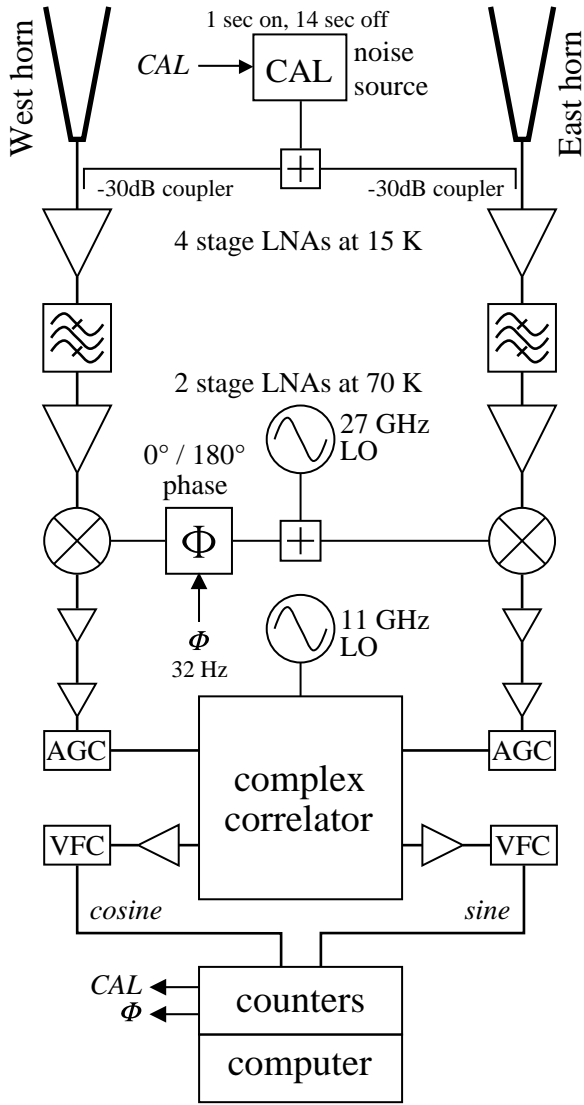}
\end{center}
\caption{\label{schem}
A schematic diagram of the interferometer. Shown from the top, down, are: 
the antennas and calibration system, the low noise receivers, the 
down--conversion system, the automatic gain control, the correlator and 
finally the logging system.}
\end{figure}

Although the receiving antennas have low sidelobes screening is placed around 
the interferometer to reduce further the likelihood of 
coherent signals entering the feed horns  from stray ground radiation.  
The screens also minimize the level of interfering RF signals.  
Fig.~\ref{schem} is a schematic diagram of the interferometer.  
A description of the various components of the system is given in the 
following sub-sections.

\subsection{The antenna system}
\label{sect:feed}

Two realizations of our antenna concept were constructed and used for 
astronomical observations.  The first, a pyramidal horn system, had an element
separation of \phl\ while the second realization, a horn-plus-reflector
system, had a \hrs\  element separation.  

\begin{figure}
\begin{center}
\leavevmode\epsfbox{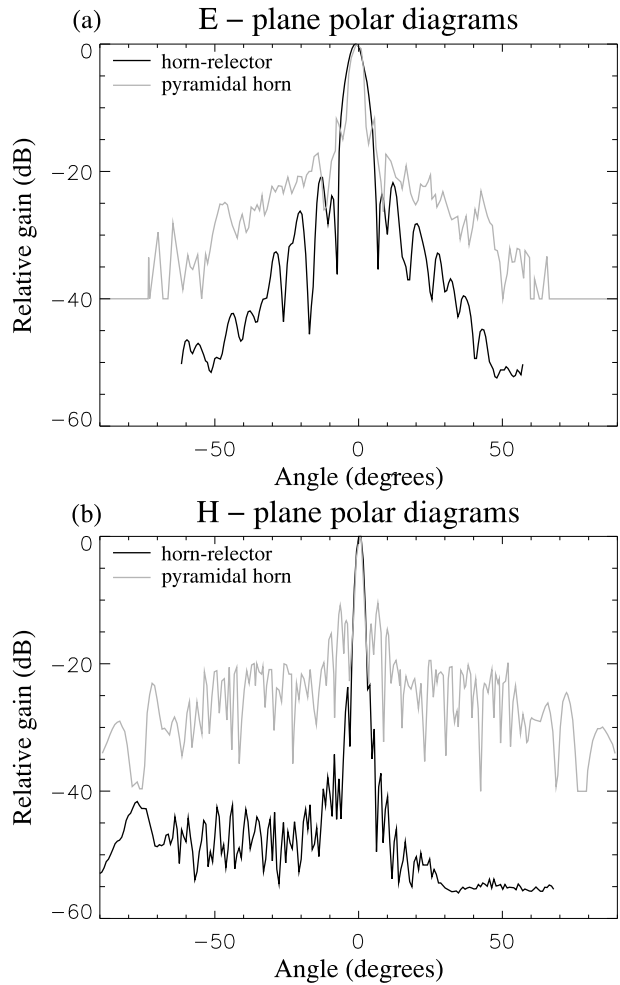}
\end{center}
\caption{\label{polar}The polar diagrams for the original pyramidal horns,
and the horn-reflector antennas. (a) shows the E-plane (horizontal) cut and
(b) shows the H-plane (vertical) cut.}
\end{figure}

(a){\em	The pyramidal horn design --- \phl spacing.} 
The E--W beam response is provided by a 
100~mm H-plane aperture while the narrower N--S beam response is achieved  by
using a 410~mm aperture across which the phase variations are corrected by  a
dielectric lens inside the aperture.  The depth of the horn is 480~mm.   The
polar diagram of each horn is $5\degr\times2\degr$ in the ${\rm E}\times{\rm H}$
 (${\rm EW}\times{\rm NS}$) directions.  The  spacing between horn centres is \phl\
(110~mm) in the EW direction to give an interferometer fringe period of
4\fdg7.
  The high side-lobe level of this design can be seen in 
Fig.~\ref{polar}.  Astronomical measurements to be described below 
indicate that the desired FWHP (full width at half power) was achieved although the aperture 
efficiency was only $\sim$0.3.

\begin{figure}
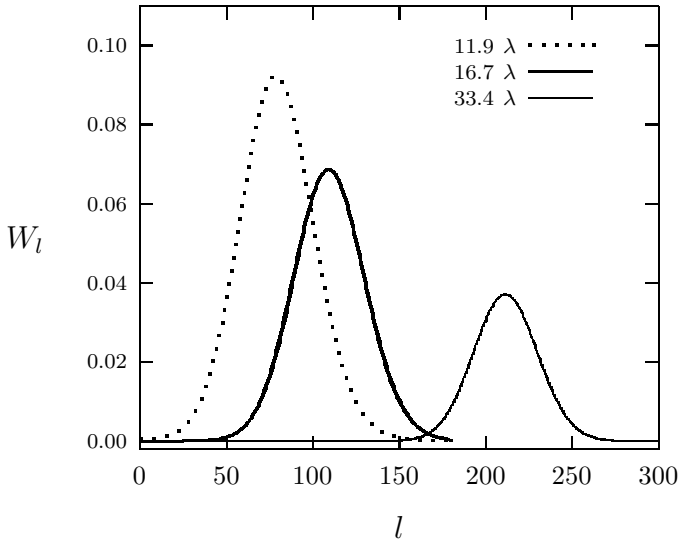

\begin{center}
\include{winplots}
\end{center}
\caption{\label{win_plots}The window functions for three antenna spacings.
The spectral power density sensitivity, $W_{l}$, is plotted as a function of
spherical harmonic, $l$. The \phl and \hrs plots are for
the pyramidal horn and horn--reflector antennas respectively.
The $33.4~\lambda$ plot shows the response for the next spacing to be used.}
\end{figure}

(b){\em	The combined horn-reflector design --- \hrs spacing.} 
 The required low side-lobe 
and high 
aperture efficiency on the desired angular scale can best be realized by 
illuminating a parabolic section reflector with a corrugated horn.  A 
detailed description will be given by Dicker, Withington \& 
Gassan (in preparation).  The physical aperture of each feed is 140 mm 
(E--W) $\times$ 400 mm (N--S) again with 
the E-plane N--S.  The FWHP is $5\degr\times2\fdg5$ (${\rm EW}\times{\rm NS}$)
and the antennas are separated by \hrs\ (152 mm) centre-to-centre.  
Fig.~\ref{polar} shows the E and H-plane polar diagrams.  A low  side-lobe
level is achieved which is consistent with a calculation that  91 per cent of
the power is within 10 dB of the beam centre, compared with  93 per cent for
a Gaussian beam profile.

The window functions of the different antenna configurations are illustrated
in Fig.~\ref{win_plots}. This shows the sensitivity of the telescope, $W_{l}$,
 to
power at different values of the spherical harmonic number, $l$,
 for \phl and \hrs spacings.
The response for a spacing of $33.4~\lambda$ is also shown, where we will be
collecting data in the future. Fig.~\ref{win_plots} shows that 
increasing the antenna separation reduces the sensitivity of the telescope,
but it should be remembered that this is largely offset by the reduced  
weather effects at larger separations.

\subsection{The RF and IF stages}
\label{sect:RF_IF}

Radio frequency (RF) signals are received by the two horns with  the E-vector
in the vertical plane.  They are then taken by waveguide  into the cryostat.
This is cooled by a Gifford-McMahon closed-cycle refrigeration  system, using
compressed gaseous helium.  The RF voltages in  each
channel of the interferometer are amplified in a 4-stage low noise  amplifier
(LNA) with a gain of 25~dB, cooled to a physical temperature of $\sim 15$~K.
A  second 2-stage amplifier, cooled to 
70~K, supplies a further 12~db of gain. When cooled
the first  amplifiers have noise temperatures between 60 and 80~K,
while the  second amplifiers have  somewhat higher noise-temperatures but
only contribute $\sim1$K to the overall noise temperature. These amplifiers
are constructed at  Jodrell Bank from Fujitsu FHR10X devices. The operating
bandwidth is  defined by a 5-pole waveguide filter between the two LNAs in 
each RF channel.  The measured RF band is 2.6~GHz wide  at  3~dB and is
centred at 32.5~GHz.

An item crucial to achieving a robust and stable operation of the 
interferometer is the calibration noise signal (CAL) which is coupled  to
each input waveguide.  The noise source is a noise diode with an excess noise
ratio of $\sim$25~dB. Its output is fed through a variable attenuator, then a
3~dB splitter into a -30~dB  cross-guide coupler on  each input waveguide to
give a calibration signal of $\sim$10~K.
  This CAL signal has 
two functions;  firstly it provides a monitor of the gain of the system 
and secondly it gives a phase reference for the interferometer. This  enables
successive days of data to be added in the correct amplitude and phase.  CAL 
is switched on for 1 sec every 15 sec.

The RF  signals from each amplifier chain are fed to a DC-biased mixer 
 where they are mixed with signals from a common Gunn diode local oscillator 
source (LO), operating at 27~GHz.  As shown in 
Fig.~\ref{schem}, the LO signal for the West channel
 has a (0\degr, 180\degr) phase 
switch which modulates the sign of the correlated output at a switch 
frequency of 32~Hz driven from the control computer.

The IF signals 
from each mixer leave the cryostat and are amplified in broad-band 
amplifiers located on a temperature stabilised plate.  Each IF channel has
four gain stages, with a total gain of $\sim$90~dB. The West IF line contains
a trombone line which can be adjusted  to make the path length the same in
the two arms of the interferometer.   This sets the interferometer on the
``white light fringe''.  An automatic gain control (AGC) system can attenuate
the IF power levels by up to 20~dB to achieve constant levels at the
correlator inputs.

\subsection{The Correlator}
\label{sect:correl}

The correlator has been designed for the wide bandwidth of 3 GHz used in 
the present interferometer.  Correlators conventionally used in long 
baseline interferometers are of the digital type so as to readily 
implement the continuous path length compensation required.  The present short 
baseline drift-scan interferometer does not require this and 
so can employ the less technically demanding analogue-analogue correlation
scheme.

\begin{figure}
\begin{center}
\leavevmode\epsfbox{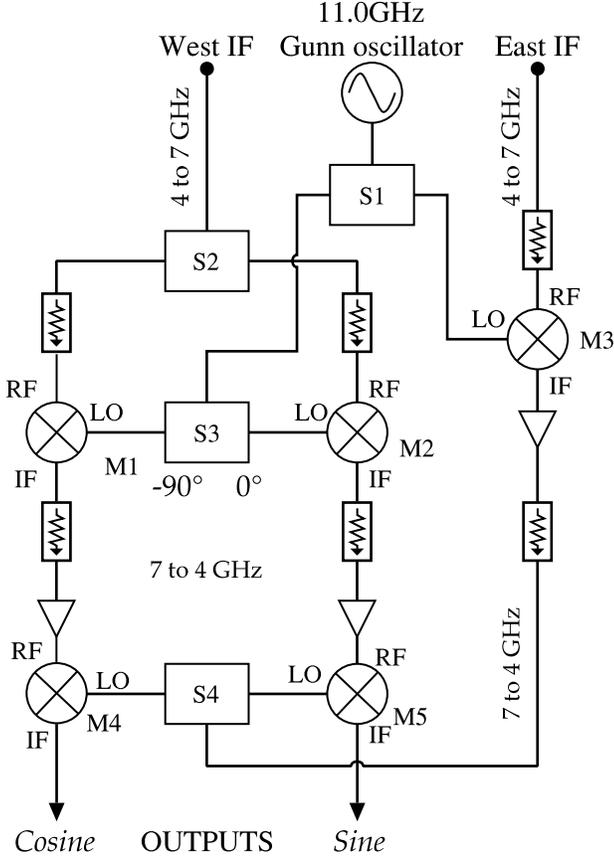}
\end{center}
\caption{\label{corr}
The correlator. Splitters are labelled S1 to S4 and the mixers are M1 to 
M5. S3 is a hybrid splitter, producing LO signals in phase-quadrature.  This
results, finally, in quadrature output signals from M4 and M5.}
\end{figure}

The block-diagram of the complex correlator with cosine and sine outputs is 
illustrated in Fig.~\ref{corr}.  Conceptually the design follows that of 
the 5~GHz short baseline interferometer which has operated successfully 
at Jodrell Bank \cite{m1997}.  However the new design is for a bandwidth of 
3~GHz rather than 0.4~GHz and the realization takes account of this much 
larger bandwidth requirement.  The orthogonal outputs are formed using
a second stage of mixing with a second LO at 11 GHz.  This LO signal is 
fed to the West IF mixers M1 and M2 via the in-phase and quadrature 
(0\degr~ and 90\degr) splitter S3, so that two West second IF signals are
produced, one 90\degr~behind the other. The East IF is mixed with an 
LO signal in M3 to give the East second IF.   Finally, the two West second IF 
signals are each mixed with the East second IF signal (at the same phase) in
mixers M4 and M5 which act as multipliers to produce cosine and sine outputs. 

The phase difference between the cosine and sine output channels was kept 
close to 90\degr~ by careful control of electrical path lengths.
This difference was found to be 85\degr~ from measurements of astronomical
signals.  The analysis  software orthogonalizes the output channels. The
efficiency of the  correlator was measured to be 0.78$\pm$0.08.

The outputs from the cosine and sine channels are applied to synchronous
voltage-to-frequency converters (SVFC) and the output pulses are sent to  the
control / logging computer.

\subsection{Control and data acquisition system}
\label{sect:control}

The computer operating the interferometer 
also performs the function of data acquisition and storage.  It keeps  an
accurate time record for registering the data by reference to a  ``Stratum
2'' time server.  A 32 Hz square-wave is  generated to supply the control
voltage for the phase switch on the  channel 1 side of the first LO.  The
same square wave drives the software phase sensitive detectors (PSD)
 which produce the cosine and sine data streams, de-modulating the 32 Hz
square waves produced by phase-switching. An  integration cycle generates a
30 sec sequence of data.  During this  cycle the CAL noise diode is switched
on twice for 1 sec each time.

The cosine and sine 16 msec data streams, and other critical parameters, are
used to produce mean and rms values for each 30 sec cycle. The rms values
provide a monitor of the performance of the system on this time scale. The
following quantities are recorded:


\begin{enumerate}
\item Time
\item PSD level: cosine and sine
\item CAL level: cosine and sine
\item rms on PSD: cosine and sine
\item rms on CAL: cosine and sine
\item First IF power: west and east channels
\item Temperature monitors within instrument enclosure.
\end{enumerate}
The local meteorology in the form of temperature and humidity from a 
Stephenson screen is recorded by the adjacent beam-switching experiments.

\begin{figure}
\begin{center}
\leavevmode\epsfbox{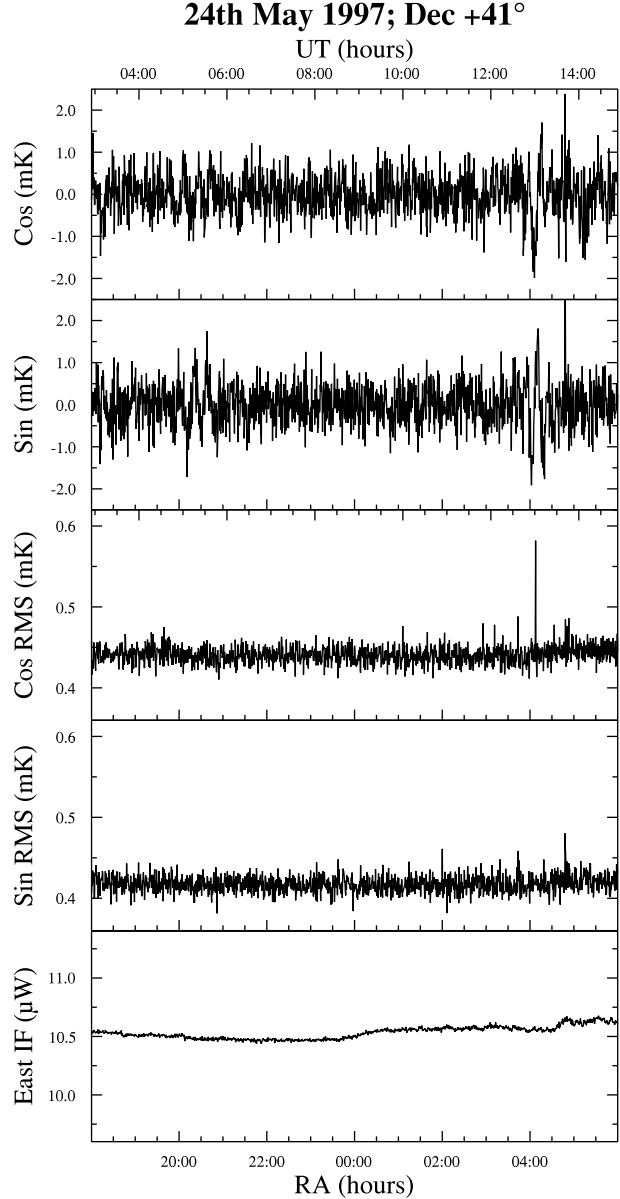}
\end{center}
\caption{\label{raw}Data from 24 May 1997.  Shown are the 
cosine and sine data channels, their errors, and the IF power in the 
east receiver.  At 4${\rm ^h}$00 RA 
fringes due to the Sun in the side-lobes can be seen.  At around 
20${\rm ^h}$00 RA,
structure from the Cygnus region is visible.  At all times the data are 
limited only by receiver noise.}
\end{figure}

Fig.~\ref{raw} is a plot of some of the data streams from 24 May 1997.
 Shown are the PSD outputs (cosine and sine), rms of the PSD outputs, and the
IF power in the east receiver. Note that the Sun is visible at UT =
13${\rm ^h}$00${\rm ^m}$ in the  side-lobes at a level $\sim$-46~dB relative
to the beam centre. The Cygnus A and Cygnus X transit is seen at  RA =
20${\rm ^h}$00${\rm ^m}$ - 21${\rm ^h}$00${\rm ^m}$ with the instrument  set
at Dec = +41\fdg0.  The rms in the 30 sec integrations is 0.48 and 0.45~mK in
the cosine and sine channels respectively.

\section{Instrumental performance}
\label{sect:performance}
\label{perf}

The interferometer was assembled at NRAL, Jodrell Bank, and tested as far 
as possible before being shipped to Tenerife for final installation in 
a 2.5 m--high aluminium-screened enclosure at Teide Observatory. 
The enclosure extends $\sim$ 1 m above the level of the interferometer horns.
This dry site 
gave the conditions appropriate for astronomical tests and calibration 
prior to the definitive astronomical programme of long-term repeated 
daily observations to achieve the deep integrations  required to detect CMB
structures.

\subsection{The interferometer response}
\label{sect:response}

We determined the beam response of the interferometer by using transits of 
the Sun and the Moon.  For angles closer than $\sim$10\degr~ to the main beam 
the response can be modelled in terms of a 
two-dimensional Gaussian modulated by a sinusoidal function in RA:
\[
\exp\left[\frac{\Delta_{1}^2}{2\sigma_{1}^2}\right]
\exp\left[\frac{\Delta_{2}^2}{2\sigma_{2}^2}\right]
\times\left\{ \begin{array}{ll}
	\cos \omega\Delta_{1}           & \mbox{\rm cosine channel} \\
	\sin (\omega\Delta_{2} + \phi) & \mbox{\rm sine channel}
\end{array}
\right.
\]
where $\Delta_{1}$ and $\Delta_{2}$ are the distances in RA and Dec
between the centre of the beam and the source;  $\sigma_{1}$ and 
$\sigma_{2}$  represent the width of the beam in RA and Dec respectively;
$\omega = 2\pi / T$ where $T$  is the separation (in the RA direction) of the
fringes on  the sky;  and $\phi$ is the offset from quadrature  of the two
channels (Section~\ref{sect:correl}).

\begin{figure}
\begin{center}
\leavevmode\epsfbox{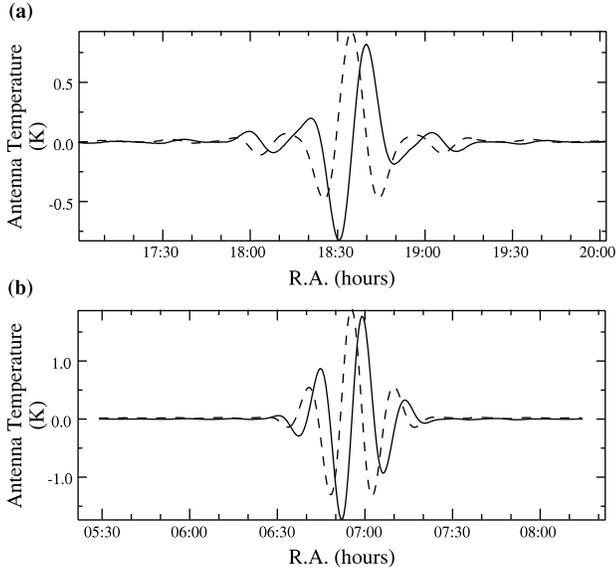}
\end{center}
\caption{\label{moon}Moon crossings from the two interferometer 
configurations. (a) is from 1 July 1996 when the  interferometer has an
antenna separation of \phl\ while (b) is from 13 April 1997 with a
separation of \hrs.  The moon semi--diameters on these dates where
16\arcmin39\farcs4 and  15\arcmin5\farcs8 respectively.  The solid line
represents the sine channel while the dashed line is the cosine channel.}
\end{figure}

Fig.~\ref{moon}(a) shows the cosine and sine channels with the 
interferometer in the \phl configuration.
    The fringe separation, corrected for the movement of the Moon,
 was measured to be 4\fdg8, and is as expected for the horn separation 
at 32.5 GHz ($\lambda = 9.2$ mm).  Although a Gaussian gives a good fit to 
the main beam response, the side-lobes become increasingly dominant at 
levels more than 20~dB below the beam peak.  As a consequence astronomical 
data were only used at times when the Sun was below the elevation of the 
aluminium screen which in practice meant night time only for this configuration.

Fig.~\ref{moon}(b) shows the Moon transit observed with 
the \hrs system.
The fringe period was measured to be 3\fdg5 as  expected for this
spacing.  The low side-lobe response shown in  Fig.~\ref{polar} was confirmed
by  transits of  the Sun in the  side-lobes, during drift scans at Dec =
+41\degr. With these low side-lobes data can be collected to within $\pm1$ hour
of local noon.

\subsection{The receiver performance}
\label{sect:RX_perf}

Hot
and cold load measurements gave a system noise estimate for both 
receivers of  120~K.  
Laboratory tests indicated that 80~K of this total 
came from the LNAs.  The atmosphere on a typical dry day contributes 
10K, mainly due to O$_2$.  We associate the additional 30~K with the 
antenna, the input waveguide system, the CAL cross-guide coupler and the
input  isolator.

The directly measured bandwidth of the receiver system was 2.6~GHz at the 
$-3$ dB (half power) points and $\sim$3.5~GHz at -10 dB.  These values were
confirmed  by measurements of the width of the  white light fringe from the
CAL signal by varying the length of the trombone line.

The rms noise recorded continuously in our CMB experiments, from the data
streams sampled at 16 msec intervals, is an effective  monitor of system
performance. The operating noise level is typically 2.2 ${\rm mK.s}^{-1/2}$.
This is 
close to the theoretical value expected for a correlator efficiency of  0.8
and a bandwidth of 2.6~GHz.  The values of system performance  for the
earlier pyramidal horn interferometer system were slightly  worse than
described above.

\subsection{Calibration}
\label{sect:calib}

The day-to-day temperature calibration of the interferometer is made through 
the noise diode signal injected into the input waveguides.  This gives 
a continuous monitor of the temperature in the correlator cosine and sine 
outputs.  In order to convert the output data to a temperature 
scale we use frequent observations of the Moon which is taken to be a 
circular disc of brightness temperature $T_b$ at 33~GHz:
\[
T_b = 214 - 36\cos (\phi - \epsilon) \:\:\: {\rm K}\nonumber
\]
where $\phi$ is the phase of the moon (measured from full moon) and 
$\epsilon = 41\degr$ is the phase offset resulting from the finite 
thermal conductivity of the Moon \cite{h1970}.
As a consequence of the dilution in the main beam of 
the interferometer, the temperature actually measured for the moon
 is comparable in size
 to the calibration signal, CAL.  
Hot and cold load measurements on the input of the interferometer give 
results consistent with the Moon and Sun (T$_b$ = 9000~K) observations.  A 
temperature calibration for CMB measurements using an astronomical 
object (the Moon) is robust since it employs the entire instrument, removing
the effects of any receiver non-linearities or sources of de-correlation in
the interferometer, and taking into account uncertainties in the efficiency
of the beam.  Such a scale can
 be directly linked to other CMB structure measurements which also 
observe the Moon.

A further astronomical check on the temperature can be made by observing 
point sources of known flux density.  The effective area of the  horn--reflector antennas gives a conversion factor of 6.46 $\mu$K/Jy.  Measurements
of the Crab Nebula (Tau A) agree with the calibration obtained using the
Moon.

\section{Atmospheric effects at 33 GH\lowercase{z}}
\label{sect:atmos}

The effect of atmospheric water vapour on the response of the 33 GHz 
interferometer has been quantified during the commissioning observations.  
A preliminary assessment has already been given by Davies et al.  (1996) of 
the atmospheric effects on the CMB observations with the 10, 15 and 33 
GHz beam-switching instruments, also located at an altitude of 2400~m 
at Teide Observatory.  As expected from a theoretical analysis 
\cite{w1994,c1995}, the atmospheric fluctuations on the smaller angular 
scales sampled by the interferometer are significantly less than those 
observed with the beam-switching experiments.  We now quantify these 
atmospheric effects observed at 33 GHz.

\subsection{The correlator and total power outputs compared}
\label{sect:int_tp}

The data streams from the 33 GHz interferometer include the total radio 
frequency power levels, measured at each first IF.  The total power  record
gives the total intensity from the sky (including water vapour  emission) in
the main beam of the interferometer.   The cosine and sine outputs give a
measure of the correlated signal in  the interferometer lobes. A comparison
can be made of the total power and interferometer outputs  over a period of
high water vapour content for each configuration of  the interferometer.	

\begin{figure}
\begin{center}
\leavevmode\epsfbox{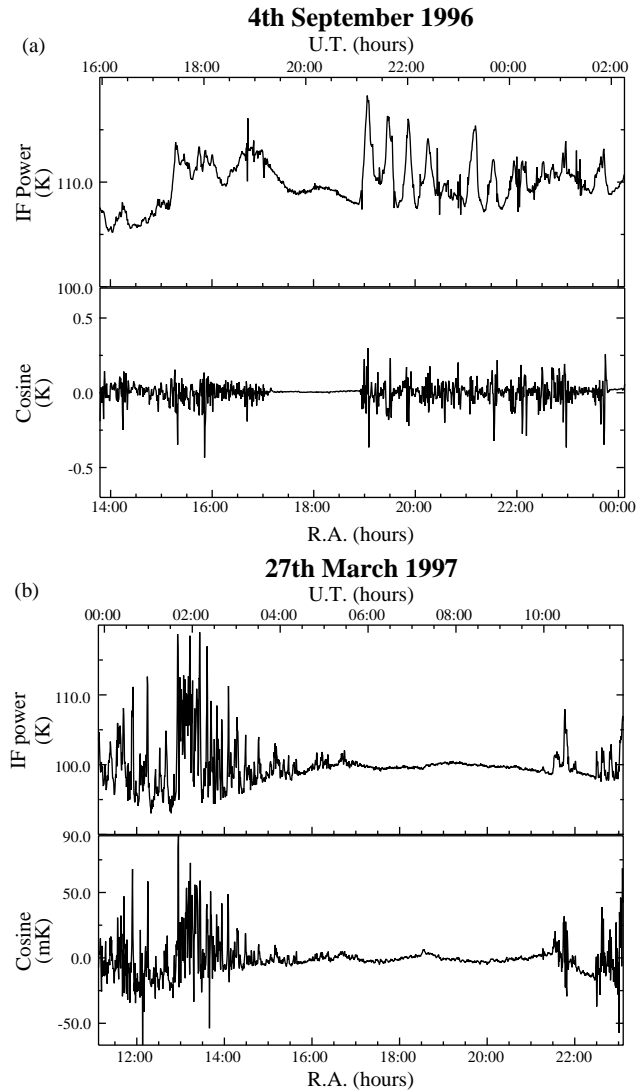}
\end{center}
\caption{\label{weather} Plots showing the rejection of atmospheric water 
vapour effects by the interferometer, on days of bad weather.  The upper part 
of each plot shows the total power fluctuations observed in the primary
 beam; the lower part of each plot shows the fluctuations on the 
interferometer cosine output. (a) \phl baseline observed on
4  September 1996 (b) \hrs baseline observed on 27 March
1997. The $\sim$10~K fluctuation in the total power records are reduced by a
factor of $\sim$100 in the short baseline data and $\sim$200 in the longer
baseline data.}
\end{figure}

Fig.~\ref{weather}(a) shows the total power and cosine records for  the night
of the 4 September 1996 with the interferometer in the  \phl
configuration.  10--20 K fluctuations in the total  power on
time-scales of 30~s are reduced to $\sim$200~mK at the cosine  and sine 
outputs. A short period of receiver noise limited data occurs around 20$\rm
^h$ UT. Fig.~\ref{weather}(b) shows a similar data  set covering the morning
of 27 March 1997 when the interferometer was in the \hrs
 configuration.  The rejection of atmospheric effects is even
more pronounced than in Fig.~\ref{weather}(a),  with the amplitude of
fluctuations in the IF power being reduced by a  factor of $\sim$200 in the
interferometer
data.

A quantitative estimate of the ratio between the interferometer output and the 
total power can be made by comparing the amplitude of the interferometer 
output calculated from the cosine and sine channels with the total power 
channel.  This ratio is $\sim$1:100 for the \phl and $\sim$1:200 
for the \hrs baselines.  The difference in ratio for the two baselines is in 
the sense predicted by Webster \shortcite{w1994}. A more detailed study of the 
atmospheric effects seen in the interferometer will be published 
separately by Watson \& Davies (in perparation).

\subsection{The interferometer compared with beam-switching}
\label{sect:int_bs}

A comparison of atmospheric fluctuations in the beam-switching and 
interferometer experiments at 33 GHz gives information about the 
angular spectrum of atmospheric structure above the Teide Observatory.  
The beam-switching experiment measures the difference between the 
temperatures seen in a 5$^\circ\times$~5\degr~ beam switched $\pm$8\degr~
 while the interferometer forms adjacent positive and negative 
lobes approximately 2$^\circ\times$~2\degr~ in size. 
Hence the angular scales sampled in  the two experiments differ by a factor
of about 4.

\begin{figure}
\begin{center}
\leavevmode\epsfbox{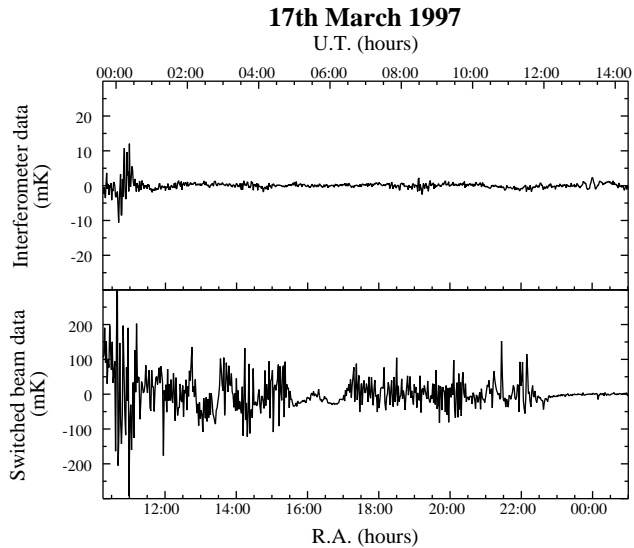}
\end{center}
\caption{\label{w2}  A comparison of atmospheric water vapour fluctuations
observed with the 33~GHz interferometer and beam-switching systems  on 17
March 1997, a day of bad weather.  The interferometer (upper part) was in the
\hrs baseline configuration, set to dec = +41\degr. The
switched beam experiment was set to dec = +35\degr.  The interferometer is 
receiver noise limited for $\sim$50 per cent of the time, and even when it is 
not the weather effects are 30 -- 50 times smaller than in the beam-switching 
experiment.	} \end{figure}

The response of the beam-switching and the interferometer systems to a 
period of strong water vapour activity experienced on 17 March 1997 is 
illustrated in Fig.\ref{w2}.  The activity was associated with 
cloud;  periods of clear sky, with a low level of water vapour,  occur before
and after the active phase.  Peaks in the beam-switching output  are several
100 mK while those in the interferometer are a few mK. 
 A quantitative comparison of the amplitude of the beam-switching and 
interferometer variations due to water vapour gives a ratio of 50:1.  
These data are for the \hrs baseline.  Some small
modification of this  value may be required to account for the difference in
declination setting  of the two experiments.  Further, the  fluctuation
amplitude is derived from 82~sec integrations in the  beam-switching
experiments and $\sim$30~sec  integrations in the interferometer.   The
effect of the latter is to increase the ratio of beam-switching to 
interferometer fluctuation amplitudes by a few tens of per cent.

\begin{figure*}
\begin{center}
\leavevmode\epsfbox{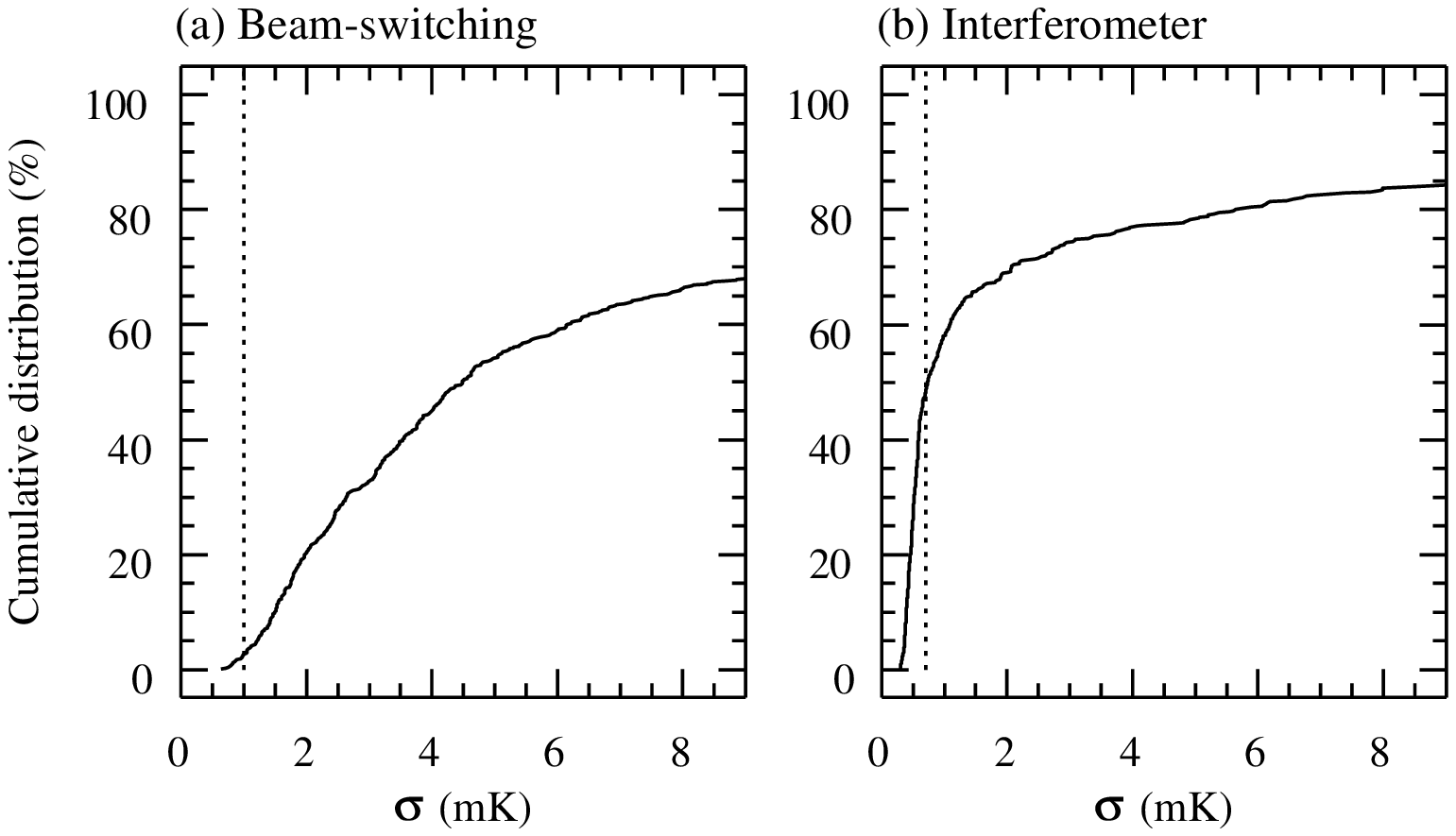}
\end{center}
\caption{\label{stat}Cumulative noise statistics for the beam-switching
radiometer (left) and interferometer (right) for September 1995, during a period
of increased water vapour activity.  The dotted lines represent the
expected rms due  to receiver noise alone. In the absense of other sources of
noise we should expect the cumulative distribution to reach approximately
65 per cent by this point. The fact that for the radiometer the percentage
is much smaller shows that it is severely affected by weather. The
interferometer is less affected.}
\end{figure*}

Because of the smaller atmospheric effects on the interferometer data, 
the fraction of useful data is significantly 
higher than for the 33 GHz beam-switching experiment.  This situation is
shown quantitatively in  Fig.\ref{stat} which is a cumulative plot of the
observed rms noise for  the raw data sets of the 33 GHz beam-switching and
interferometer  experiments.  The plots are for night-time data taken during
 September 1995, a period when the water vapour content begins to rise 
following the drier summer months.  Data for the plots were the mean rms 
signals, examined in 30${\rm ^m}$~ and 12${\rm ^m}$~ RA intervals for the 
beam-switching and interferometer 
data respectively;  these intervals correspond approximately to the 
angular scale of maximum sensitivity for sky signals in each experiment.  
We see that the fraction of data with an rms less than that expected for
receiver noise alone is a few 
per cent for the beam-switching and $\sim$50 per cent  for the 
interferometer.  Hence the time for which observing 
conditions are suitable for interferometry at 33 GHz is many times greater 
than for beam-switching observations. In a one year period the availability
of the interferometer is $\sim 70 - 80$ per cent, compared with $\sim 20$
per cent for the radiometer. This topic will be examined in more detail in
the paper by Watson \& Davies (in preparation).

\section{Astronomical tests of the interferometer}
\label{sect:astro}

In this section we describe astronomical tests which demonstrate that the 
interferometer has the necessary sensitivity and stability to detect 
CMB fluctuations with $\Delta T_{\rm rms}\:\sim$50~$\mu$K when a realizable 
amount of data are stacked.  Any strong astronomical source
 can be clearly detected when data from a number of transits are stacked.   In
the case of the strong sources, Tau A, Cyg A, Cyg~X and Cas A,
 approximately 10 transits 
are sufficient.  The normal observing procedure is to take RA drift scans 
of 24${\rm ^h}$~ duration with the interferometer set at a fixed declination.  
This technique ensures that any correlated pick-up from the ground 
remains fixed, thereby contributing to the stability of the baselines.

With the \phl baseline (pyramidal horns) only night-time 
data could be used, while with the \hrs baseline  
(horn + reflector design) data could be taken up to within $\pm1$ hour 
of the Sun transit.

\subsection{Dec = +41\degr~ scans}
\label{sect:dec41}

\begin{figure*}
\begin{center}
\leavevmode\epsfbox{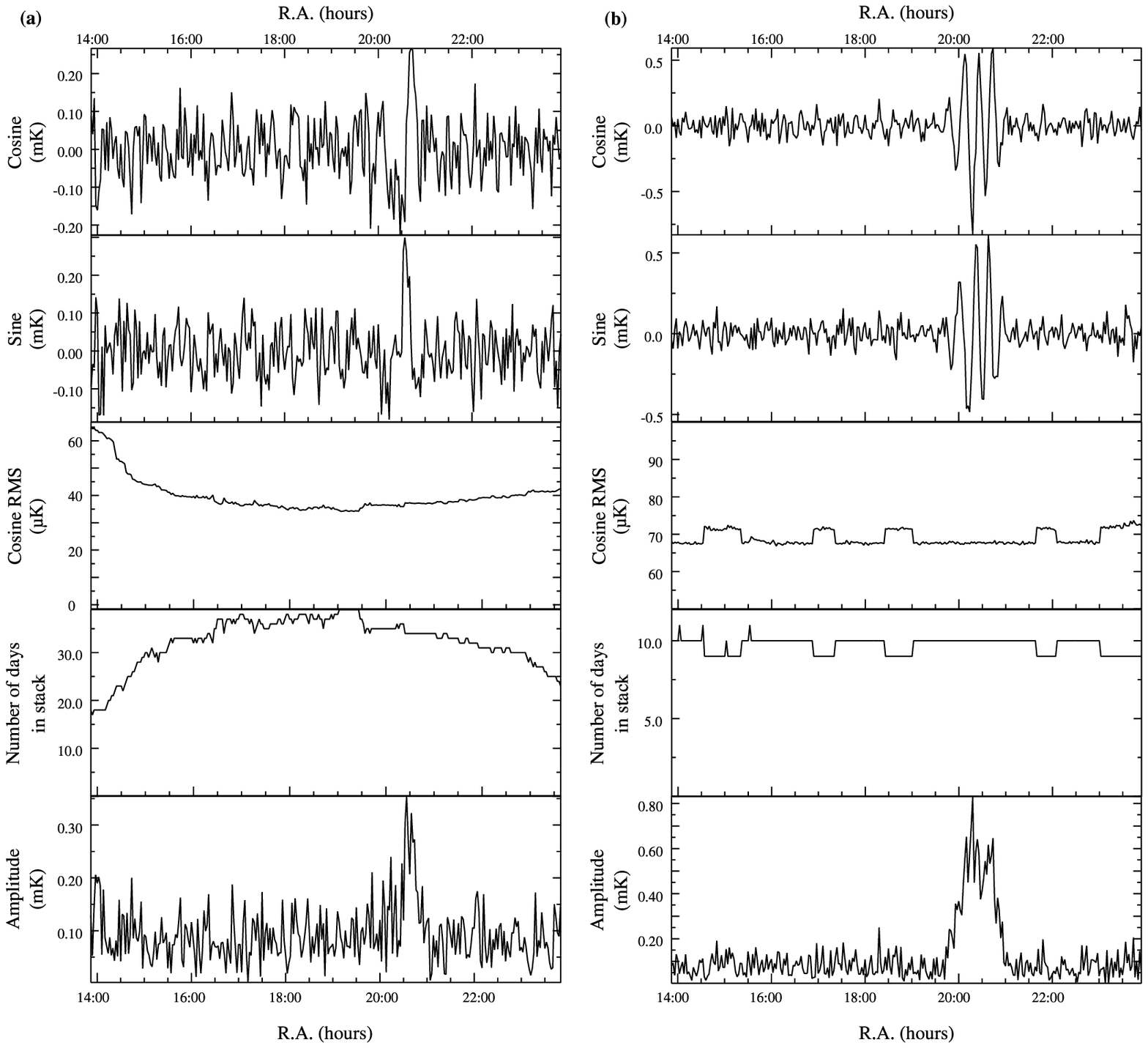}
\end{center}
\caption{\label{41}Stacks of data taken at Dec = +41\degr; (a) Data taken 
with the \phl antenna spacing; (b) Data taken 
with the \hrs antenna spacing.  The data were taken over 
October/November 1995 and April 1997 respectively. Only the rms 
values for the cosine channel are shown as they are very similar to the  sine
rms values. Each point represents an integration over 2 minutes of RA.}
\end{figure*}

This declination was chosen for observation because it lies in the centre 
of the region that we have been using for our deep surveys of CMB
fluctuations  \cite{d1996,h1997} and for the present purposes it  contains
the extragalactic radio source Cyg A (RA = 19${\rm ^h}$59${\rm ^m}$,  Dec =
40\degr 44\arcmin, J2000) and the Galactic HII complex Cyg X \cite{p1964} which  covers an
area bounded by RA = 20${\rm^h15^m}$ to 20${\rm^h45^m}$  and Dec = 38\fdg5~
to 44\fdg0.   Fig.~\ref{41} shows the cosine and sine plots of the Dec =
+41\degr~ stacks  taken with the \phl and \hrs
 baseline configurations.  The 
complex structure in Cyg X gives different responses in the two 
interferometer configurations as expected.  The longer baseline just 
separates Cyg A (flux density S = 35$\pm$7 Jy) from Cyg X at the  expected
temperature of 230 $\mu$K for the sensitivity of T = 6.46 $\mu$K  for 1 Jy. 
Away from Cyg A and X the data are noise-limited with  $\Delta T_{\rm rms} 
\sim$50 $\mu$K in 2~min integrations stacked over approximately 10 days.  
This is the value expected for the measured system noise and bandwidth with 
a correlator efficiency of $\sim 0.8$.  If the performance improves as
expected,  on further stacking and data processing an rms of 6  $\mu$K per
beam would result from the addition of 100 scans.  The data must then be 
corrected for the dilution of the sky signals in the interferometer lobes 
compared with the primary beam envelope. The size of this correction  depends
strongly on scale size, and the model assumed for the CMB power spectrum and
will be discussed in later papers.

\subsection{Observations of Cas A and Sgr A}
\label{sect:CasA_SgrA}

\begin{figure*}
\begin{center}
\leavevmode\epsfbox{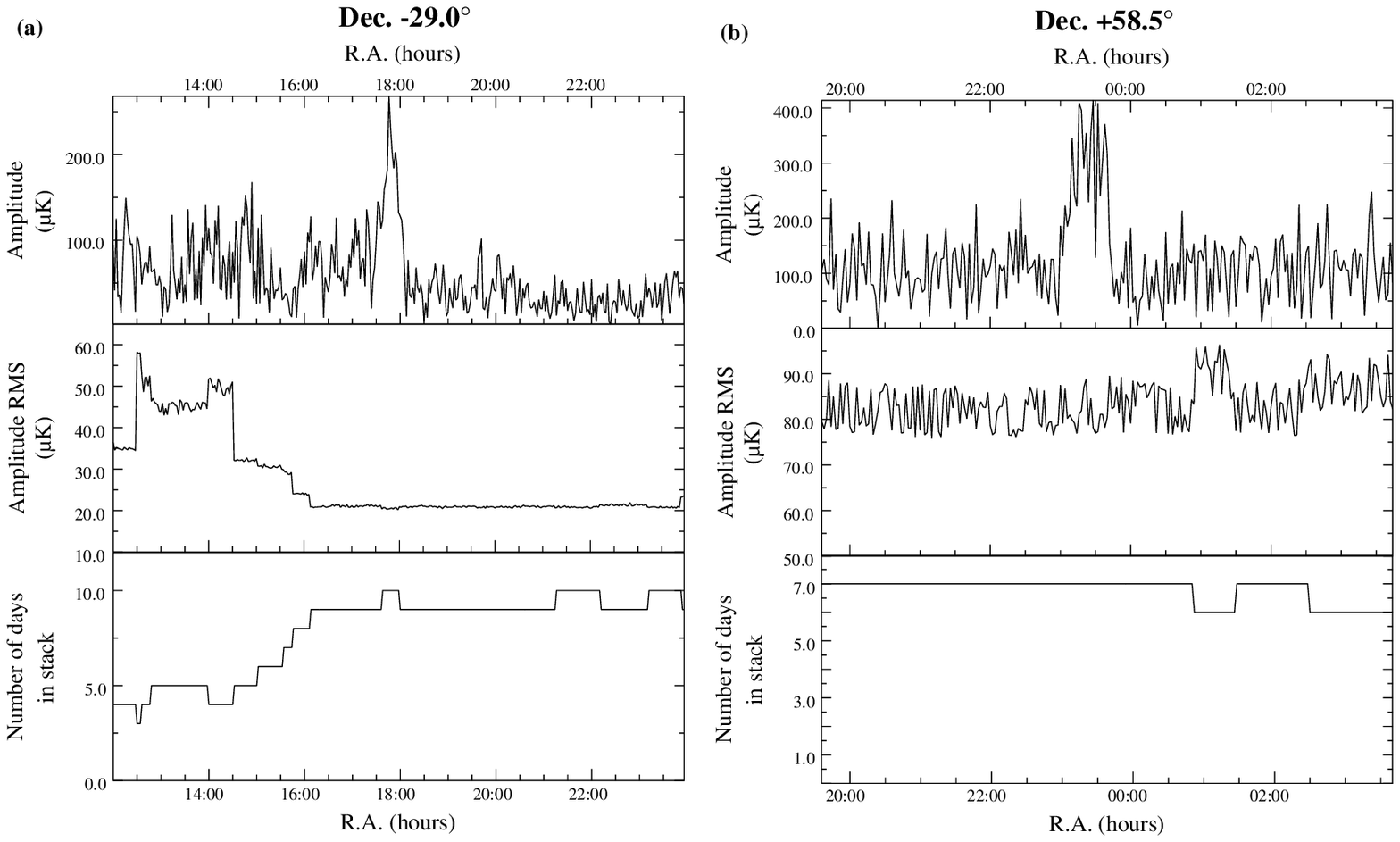}
\end{center}
\caption{Sgr A (left) and Cas A (right) as seen by the 
interferometer with a \phl spacing.  For each source, the amplitude 
$(\sin^2 + \cos^2)^{1/2}$, the rms error of this, and the number of 
days of observations at each RA, are shown.  All these observations 
were taken over September -- November, 1996}
\label{fig:SgrA_CasA}
\end{figure*}

Fig.~\ref{fig:SgrA_CasA} shows the amplitude and rms noise plots for Cas A
and Sgr A.   The interferometer was set at the declination of Cas A  (RA =
23${\rm ^h}$23${\rm ^m}$,  Dec = 58\degr49\arcmin, J2000) and 7 scans from September 1996
were added to  give the plot in Fig.~\ref{fig:SgrA_CasA}(b).   The dimensions
of Cas A  ($6\arcmin \times 6\arcmin$) make it unresolved by  the
interferometer beam. The flux density expected at 33 GHz,  taking account of
its secular decrease in intensity \cite{baars} is  162$\pm$19 Jy at the
present epoch so the expected antenna temperature is  $\sim$400~$\mu$K for
the pyramidal horn antennas.  This is the observed  value. 

The stack of 10 scans through Sgr A (RA = 17${\rm ^h}$46${\rm ^m}$ Dec = 
$-29^\circ$00\arcmin, J2000) 
is shown in Fig.~\ref{fig:SgrA_CasA}(a).  A signal amplitude of 250 $\mu$K is 
observed.  For the pyramidal horn configuration this temperature corresponds 
to a flux density of $\sim$120 Jy.  Little published information is available 
on the 33~GHz brightness distribution in the Galactic centre region.   The
central $2\degr\times2\degr$ contains both synchrotron and free-free 
emission on a range of angular scales \cite{m1979}. The observed flux in
the interferometer lobes is  not unreasonable.

Both the Cas A and Sgr A stacks give $\Delta T_{\rm rms} \sim$50 $\mu$K in a 2 
min integration for a stack containing $\sim$10 scans.  This is consistent 
with the Dec = +41\degr~ scans and indicates that the performance of the 
interferometer is not being affected by the weather in any subtle low-level 
way since these observations were made at various times throughout the year
and at various elevation angles (declinations).

\section{Conclusions}
\label{sect:conc}

We have demonstrated the potential of interferometric CMB observations at 
33 GHz from the Teide Observatory in Tenerife, at an altitude of 2400~m.  
This instrument will extend to 2\degr~ the angular range of CMB fluctuations 
observed in the Tenerife CMB project which at present covers the range 
5\degr~ to 15\degr.  We describe the design  and performance of this
broad-band,  short-baseline interferometer and demonstrate its phase
stability and  sensitivity with observations of Cyg A, Cyg X, Cas A and Sgr
A.

The interferometer shows an improvement in atmospheric fluctuation 
rejection over the 33GHz  beam-switching radiometer, which has its
peak sensitivity on 8\degr~ scales, by a  factor
greater than 30.  This allows the interferometer to observe for 70 -- 80 per
cent observing  efficiency on the site at 2\degr~ scales and vindicates the
choice of  Tenerife for interferometry and the Very Small Array \cite{j1998}.

The first observations with the 2\degr~ interferometer will be a deep scan at 
Dec = +41\degr~ to explore the  CMB fluctuation field at high Galactic
latitudes.   This area of the sky has been covered with a deep survey at 10,
15 and  33 GHz which has already shown that the Galactic foreground
contribution  is smaller ($<$10 per cent) than the CMB fluctuation level
at 33 GHz on  8\degr~ scales;  the Galactic contribution will be
significantly less  than this on the smaller, 2\degr, scale of the 
interferometer \cite{dw1998}.

Based on the observations reported here for the horn-reflector interferometer 
we estimate that the noise $\Delta T_{\rm rms}$, on the antenna temperature,
for a stack of 100  scans will be 20 $\mu$K in a 2~min interval in RA.  By
combining data from over the  $\sim$20 minute  interferometer response
pattern, this becomes $\sim$6 $\mu$K per beam.  On conversion  to sky
temperature this becomes 20 $\mu$K per beam.  We expect to cover at  least
two such 24${\rm ^h}$~ RA strips per year and to generate a sky map with 
$\sim$500 beam areas; this will result in an error of $5 - 10$ per cent from
cosmic sample variance considerations. We believe that the calibration error will
be less than 10 per cent. Based on these estimates we expect an accuracy in
determining the $\sim 50~\mu{\rm K}$ CMB signal on a 2\degr~scale of $\sim 10$
per cent.  By changing 
the antenna spacing it is possible to  obtain data covering an angular scale
$1\degr - 2\degr~$ which will enable us to explore the rise to the first Doppler
peak.

\section*{Acknowledgments}

This work has been supported by the European Community Science programme 
contract SCI-ST920830, the Human Capital and Mobility contract 
CHRXCT920079, and the UK Particle Physics and Astronomy Research Council.
We thank the technical staff at Jodrell Bank who have made a major 
contribution to the successful construction of the interferometer; in 
particular John Hopkins, Neil Roddis and Colin Baines.

\label{lastpage}

\end{document}